\documentclass[aps,twocolumn,epsfig,showpacs]{revtex4}
\usepackage{graphicx}
\usepackage{epsfig}
\begin{document}
\title{Force induced melting of the constrained DNA} 
\author{Amit Raj Singh, D. Giri$^\dagger$ and S. Kumar} 
\affiliation{Department of Physics, Banaras Hindu University,
     Varanasi 221 005, India \\
$^\dagger$ Department of Applied Physics, Institute of Technology, Banaras Hindu University,
Varanasi 221 005, India} 

\date{\today}
\begin{abstract}
We develop a simple model to study the effects of an applied force on the 
melting of a double stranded DNA (dsDNA). Using this model, we could study  the 
stretching, unzipping, rupture and slippage like transition in a dsDNA.
 We show that in absence of an applied force, the melting temperature and 
the melting profile of dsDNA strongly depend on the constrained imposed on 
the ends of dsDNA. The nature of the phase boundary which separates the 
zipped and the open state for the shearing like transition is remarkably 
different than the DNA unzipping.
\end{abstract}
\pacs{36.20.Ey, 64.90.+b, 82.35.Jk, 87.14.Gg}
\maketitle

\section{INTRODUCTION}
Properties related to structure, functions, stability etc. of the 
bio-molecule are the results of inter and intra molecular forces 
present in the system \cite{albert,israel}. So far  the measurement of these 
forces were possible
through the indirect physical and thermodynamic measurements like 
crystallography, light scattering and nuclear magnetic resonance spectroscopy 
etc. \cite{Wartel_Phys.Rep85}.
For the direct measurement of these forces, it is essential that the state
of the system be monitored while an independent force is applied 
\cite{Smith_Science92,Lee_Science94,Rau_BPJ92,Boland_PNAS92,Noy_ChemBIO97}.
In recent years single molecule force spectroscopy (SMFS) techniques 
such as optical tweezers, magnetic tweezers, atomic force microscope (AFM)  etc 
have measured these forces directly and many important information 
about the bio-molecules have been inferred \cite{Smith,Schu,Lavery,busta,Gosse,busat1}.
 Now it has also been  realized that the measurement of these forces not 
only depend on
the molecular interactions present in the system but also on the loading rate, direction
of the applied force \cite{Lee_Science94,Strunge,Cludia} etc. Moreover, 
these experiments also provide a platform where various theoretical models 
and their predictions can be verified.

In this context considerable efforts have been made to study the 
separation of a double stranded DNA (dsDNA) to two single stranded 
DNA (ssDNA). 
Understanding the mechanism involved in separation of dsDNA may shed 
light on  the processes 
like transcription and replication of DNA \cite{albert,Baker}. At equilibrium, 
DNA will separate when the free energy of the separated ssDNA
is lower than that of the dsDNA \cite{freenergy}. In most of the biochemical studies of 
DNA separation, the strands separate upon increasing the temperature (T)
of the sample until the DNA melts (DNA melting or thermal denaturation). 
However, in {\it vivo},
DNA separation is not thermally driven, rather mediated by enzymes and other 
proteins \cite{albert,Bockelmann}. Mechanical separation of dsDNA  using SMFS techniques is known
as DNA Unzipping ( Figs. 1a and 1b) at temperatures, where the dsDNAs 
are stable, have recently been performed. The force (f) required to break a 
base pair is about
15 pN \cite{Bockelmann, Bock}. A large number of theoretical and numerical 
efforts \cite{Somen,Sebastian,Lubensky, Marenduzzo,kgbhat,eye} have 
been made 
to gain further insight into the mechanism of DNA opening. One of the major 
result from these  studies was the prediction of re-entrance in the low 
temperature region. \cite{kgbhat,eye,Somen1,Marenduzzo1}

\begin{figure}[h]
\includegraphics[width=3.0in]{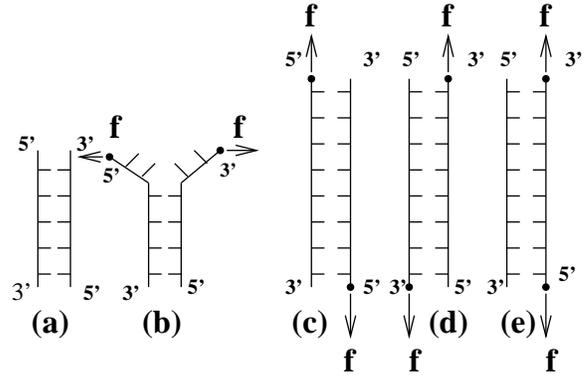}
\caption{Schematic representation of dsDNA: (a) dsDNA in zipped form; 
(b) Unzipping of dsDNA by the force ($f$) applied at one end $(5'-3')$; 
(c and d) Shearing by the force along the chain applied at the opposite ends 
$(5'-5'$ or $3'-3')$ of the dsDNA; Fig. (e) represents the case where the force 
has been applied at $5'-3'$ end of the same strand of the dsDNA.}
\label{fig-1}
\end{figure}

In some cases, the term unzipping is not appropriate because the inter chain
interactions may be carried in a different way. For example, instead of pulling 
a chain of opposite strands at 5' and 3' (Fig. 1b), it is possible to pull 
the chain on the opposite ends of two strands at 5' and 5' end (Fig. 1c) or 
3' and 3' ends (Fig. 1d) or 5' and 3' ends of the same strand (Fig. 1e). 
It is found that in these cases [figs. 1(c) and 1(d)], transition is akin to 
shearing like. The unbinding force strongly depends on the pulling end and lie 
in between  50-150 pN \cite{Strunge,Lee_Science94,Cludia} which is much larger 
than the unzipping force.

The aim of this paper is to understand the effect of pulling force  
DNA melting under the various constrain imposed on the ends of dsDNA. 
In section II we develop the model and discuss two methods namely 
the thermodynamic analysis and exact enumeration technique to study the 
force induced melting of dsDNA. The nature of the phase boundary near
$T=0$ and the limitation of the analysis will be discussed in this 
Section III comprises results obtained for DNA unzipping, dissociation of dsDNA 
and the effect of bulge movement in dsDNA. The paper ends with brief discussion 
in section IV.

\section{MODEL AND METHOD}

\begin{figure}[th]
\includegraphics[width=2.5in]{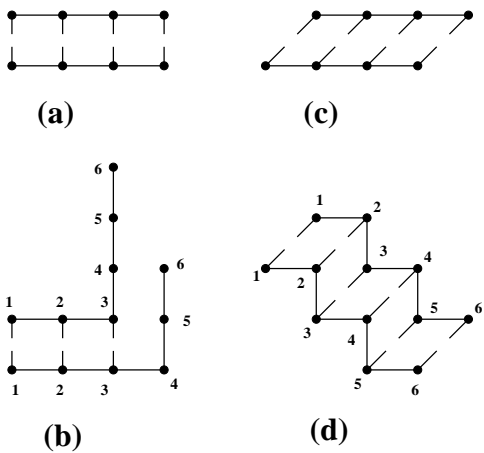}
\caption{ (a \& b) are the schematic representations of some of the 
conformations of the model introduced in Ref. \cite{kgbhat,eye,Grass,jcp}. In this
model, we have  only one ground state conformation [Fig. 1(a)]. Because of lattice 
constrains, other conformations of zipped state are not possible. (c \& d) 
are schematic representation of  dsDNA conformations with diagonal interaction
which leads to the large number of conformations of the zipped state. 
}
\label{fig-2}
\end{figure}

We consider two linear polymer chains which are mutually-attracting-self-avoiding 
walks (MASAW's) on a square lattice as shown in Fig. \ref{fig-2}. This is 
the simplest  model of dsDNA where $i$-th monomer of one strand 
can interact with the $i$-th monomer of other strand only \cite{kgbhat,eye,Amit,Grass,jcp}. This kind 
of base pairing interaction is similar to the one studied in Poland Scheraga 
(PS) model or Peyrard Bishop (PB) model \cite{polsch,polsch1}. However, 
in the present model 
configurational entropy of the system has been taken explicitly which is 
ignored in these (PS or PB) models. In order to study the response of an applied 
force on melting, we consider following cases as discussed above: (I) a pulling force 
may be applied on the chain at the 5'- 3'end (Fig. 1(b)). This will correspond
to the situation of DNA unzipping. (II) For the slippage, a force may be 
applied  along the 
chain at two opposite ends of the dsDNA {\it e.g} 5'-5'end (Fig. 1(c)) or 3'-3'end 
(Fig.1(d)). 
Two interesting situations may arise for slippage: (a) if pulling is fast, at some 
critical force $f_c$, the rupture occurs and dsDNA dissociates to the two 
single strands of DNA (Fig. 3(b)) \cite{Lee_Science94}. 
In this case the system has a larger energy barrier for the complete unbinding. 
The other possibility involves the slow pulling, where small bulge loops  can 
form 
in the chain and propagate to the pulling end (Fig. 3(c-e)). This process 
requires spontaneously binding and unbinding of few bases and through the 
process of diffusion, a bulge slides over the other chain with small 
energetic barrier \cite{Neher}. 

Neher and Gerland theoretically studied the 
dynamics for force induced DNA slippage \cite{Neher} for the homo-sequence (bulge movement) 
and hetero-sequence (dissociation of two strands) and found the expression for 
the critical force. However, in their studies, they have
also ignored the configuartional entropy of the chain and hence provide
a limited picture of the mechanical separation of dsDNA.

\subsection{Thermodynamics of force induced melting}

The thermodynamic of force induced DNA melting can be obtained from
the following relation \cite{bloom}
\begin{equation}
\Delta G= \Delta H -T \Delta S -f.x
\end{equation}

where $G$, $H$, $S$ and $x$ are the free energy, enthalpy,  entropy and 
reaction coordinate (end-to-end distance in this case) of the system
respectively. To determine the nature of phase boundary, we put $\Delta G =0$
which gives
\begin{equation}
f x = -\Delta H -T\Delta S 
\end{equation}

The entropy defined in the Eq. 1  has contributions from the  configuartional
entropy of the zipped DNA ($S_z$), entropy associated with the open state
($S_o$) and entropy associated with dissociated chains ($S_u$) etc.
In unzipping the applied force does not influence the entropy of the chain
while in shearing, it does. For the unzipping, we can write
\begin{equation}
f x = -\epsilon N' + N'TS_z - 2 (N-N')T S_{o}  
\end{equation}
where $\epsilon$ is the effective base pairing energy. 
At low temperature {\it i.e.} near $T = 0$, all bases will be intact ($N'= N$) and
hence there will be no contribution from the open conformations. Moreover, 
the second term
in Eq. 2 stabilizes the zipped state.  Eq. 3 may be written as

\begin{equation}
2 f N = -\epsilon N + NTS_z   
\end{equation}

The factor 2 comes from the fact that chain is in unzipped state and the
distance between the extreme ends is equal to $2 N$ . We substitute the value
of $\epsilon =-1$ in Eq. 4 which gives
\begin{equation}
f= 0.5+ \frac{1}{2} T S_z
\end{equation}

This is in accordance with earlier studies \cite{Somen,Marenduzzo,kgbhat,eye} 
that the applied 
force increases with the temperature at low temperature which is a signature 
of re-entrance. At higher $T$, the chain will start opening and the third 
term of Eq. 3 associated with open state will start cooperating with the
applied force and hence the applied force start decreasing after certain 
value of temperature.

Unlike unzipping, in case of shearing (rupture or slippage), the applied
force competes with the entropy associated with the zipped configurations
so that the chain acquires first the stretched state and then start opening.
In such a situation entropic contribution of zipped chain (second term of 
Eq. 3) at the phase bounadry will be  absent. However, there will be an 
additional contribution of entropy associated with the unzipped chain. 
At low temperature for the rupture ($x = 1)$, we can write

\begin{equation}
f  = -\epsilon N' - 2 N T S_u + 2 (N - N')T S_{o}  
\end{equation}

At $T = 0$, Eq. 6  gives the force required for rupture which is
equal to $N$.  Up to certain temperature when the intact bases remain 
equal to $N$ the entropy associated with open conformation will be  
zero and hence the expression for the applied force (rupture) can be 
written as

\begin{equation}
f   = N - 2 N T S_u   
\end{equation}

Above this temperature,  $N'$ decreases with
temperature and hence bubble forms, therefore, more force is needed to
keep system in the stretched state. Therefore, the phase boundary between
zipped and open states should bend. For shearing like transition, $x = N$ and
hence required force is equal to $1$ and should have similar behavior.

The precise value of entropic contribution  near the phase boundary
is difficult to obtain analytically. Therefore, it is not possible to get 
the entire phase boundary from the Eqs. 3 $\&$ 6. Using the exact enumeration
technique \cite{vander}, contribution of $S_o$ may be obtained for the finite size 
chain and an estimate of the phase boundary may be obtained.

\subsection{Exact Enumeration Analysis}

The unzipping case for the model proposed above has been studied in detail
\cite{kgbhat,eye,jcp}. It was shown that force temperature diagram demarcates 
the zipped and unzipped state and unzipping force decreases with temperature
without any re-entrance. The absence of re-entrance in force-temperature plain
is due to the ground state entropy of the zipped state which has been 
suppressed because of the imposed lattice constraint on the base pairing 
interaction (Fig. 2 (b)). However instead of base pairing interaction 
taken in Ref. \cite{kgbhat,eye,jcp}, if one considers the diagonal interaction
shown in Fig. 2(c) and 2(d), one may observe the re-entrance in the proposed model 
also. The choice of diagonal interaction is analogous to the walks on the 
oriented square lattice \cite{somen_prl}.

The model presented here may describe above mentioned effects provided we also
incorporate effect of movement of bulge in the partition function. For the 
unzipping, we fix one end of the dsDNA and apply force at other ends (5'-3') as 
shown in Fig. 1b. In order to study the behavior of slippage,  we apply force 
at opposite ends (5'-5' or 3'-3') of the strands.  We model the fast pulling 
(i.e. dissociation of two strands), by not allowing the formation of base pair in 
the model after the chain slides over the other strand.
However, for the diffusion of bulge in homo-sequence (slow pulling), we 
apply a force on the opposite strands (5'-5') so that chain 
acquires the stretched state. If force exceeds further, the chain moves one unit 
towards the applied force direction (Fig. 3c-e). 
Since spontaneous binding and unbinding is possible, now we allow the 
formation of base pairing of $(i+1)$-th base of one chain to ith base of other 
chain (Fig. 3e) and calculate the partition function ($Z^{(1)}$) of the re-annealed chain. 
For the 
next unit of displacement, we allow $(i+2)$-th base to interact with the 
ith base and calculate the partition function ($Z^{(2)}$) and so on. 
In this way, we can construct a series of partition functions ($Z^{(i)}$) for 
the slippage. 
It may be noted that
for the unzipping case  we monitor the displacement $x$ along the force 
direction while for the slippage case, we monitor the displacement $y$ along the force 
direction.  The contribution to energy due to this force, $f$, is $-f x$ (or $-f y$).

\begin{figure}[t]
\includegraphics[width=2.5in]{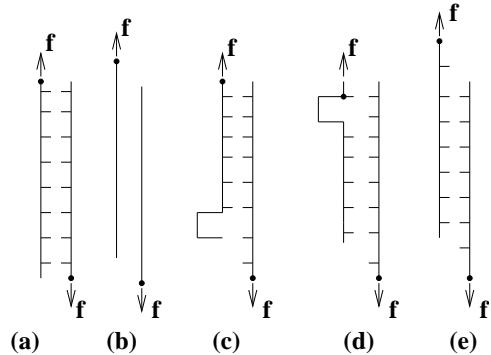}
\caption{ Schematic representation of slippage of DNA: (a) dsDNA in complete stretched 
form under the application of force; (b) Dissociation of dsDNA in two single strand 
DNA (ssDNA) at some critical force $f_c$ without any base pairing; Figures (c-e) show the 
schematic representations of bulge movement along the chain. Even if one of the chain 
slides over the other, the base pairing between $i$-th nucleotide of one strand with 
$(i+1)$-th nucleotide of the other strand is possible.}
\label{fig-3}
\end{figure}

We enumerate all conformations of MASAWs whose one end is fixed and other
end is attached with the pulling device (e.g. tip of the AFM). 
We specifically monitor the reaction coordinate {\it i.e.} 
end to end distance or distance between the fixed end and tip of 
the AFM.  The partition function of the system under consideration can
be written as a sum over all possible conformations of dsDNA:
\begin{eqnarray}
Z_N & = & \sum_{all \; walks}^{N} {x_1}^N {x_2}^N \omega^m u^x \nonumber \\
    & = & \sum_{m,x} C(m,x) {x_1}^N {x_2}^N \omega^m u^x
\end{eqnarray} 

where $N$ is the chain length ({\it i.e} $N$ steps walks) of each strand consisting 
of $N$ bases. $x_1$ and $x_2$ are the fugacities associated with each 
step of the two self-avoiding walks representing the two strands respectively. 
({\it For simplicity we take  $x_1 = x_2 = 1$ for our calculation}).
$\omega$  (= $\exp (-\beta \epsilon)$ is the Boltzmann weight associated with 
the binding energy ($\epsilon$) of each diagonal nearest neighbor pair and $m$ 
is the total number of such pairs in the chain. 
$u$ ( = $ \exp[\beta (\vec{f}.\hat{x})]$)  ($\hat x =$ unit vector along $x-$axis) 
is the Boltzmann weight associated with the force. 
$C(m,x)$ is the number of distinct walks of length {\bf $2N$} having $m$ number of  pairs 
whose end points are at a distance $x$ apart. 
We have obtained $C(m,x)$ for $N \le 15 $ bases and  analyzed the partition 
functions. 

Quantities of interest like reaction coordinate ($x$ or $y$) 
and fraction of base pairs can be calculated from the following 
expressions:

\begin{equation}
<x>=\frac{\sum x C(m,x) \omega^m u^x}{\sum C(m,x) \omega^m u^x}
\end{equation}
\begin{equation}
<m>=\frac{\sum m C(m,x) \omega^m u^x}{\sum C(m,x) \omega^m u^x}.
\end{equation}

\begin{figure}[t]
\includegraphics[width=3.in]{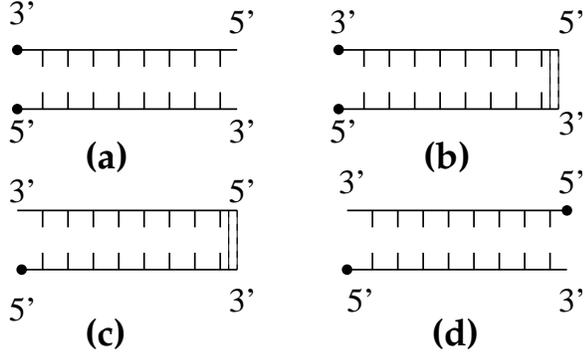}
\caption{Schematic representation of the various confinements imposed on 
the end of dsDNA in absence of force: (a) One end (5'-3') of both strand 
is fixed while other ends are free; (b) One end (5'-3') of both strand is kept 
fixed and other ends (3'-5') of both strands are tied together. In this case chain 
opens from the middle; (c) Same as (b), but here only one end (5') of one strand 
of dsDNA is kept fixed; (d) It represents the complete zipped-stretched  
state where 5'-5' ends are kept fixed while 3'-3' end are free. For all these 
cases melting profile depends on the constraints imposed on the end of the
strands shown by black circles. 
}
\label{fig-4}
\end{figure}

Since dissociation of dsDNA and bulge movement  are dynamic phenomena, 
which can be considered  in a quasi-static equilibrium. Moreover, we monitor 
the distance of the end points of the dsDNA where the force has been 
applied. In view of above, we do our analysis in constant 
distance ensemble (CDE) where temperature has been kept constant. 
The partition function in {\bf CDE} may be defined 
as $Z_N(x,T) = \sum_m \exp(\beta m \epsilon)$. The two ensembles are related by
$Z_N(T,F) = \sum_x Z_N(x,T) \exp(\beta f x)$.
The free energy is given by the relation $F_N(x,T)= -T \ln Z_N(x,T)$ and 
average force $<f_c>$ is thus $\frac{dF}{dx}$.
It is pertinent to mention here that in {\bf CFE} the average separation 
$<x>$ fluctuates while in {\bf CDE} one measures the average force to keep 
the separation constant at a fixed temperature.

Since most of the single molecule experiments are performed for 
finite size chain and the fact that no true phase transition can occur 
in single molecule, we consider only finite chain calculation and calculate  
the ``state diagram". The boundary of state diagram (F-T diagram) can be 
obtained from the maxima of fluctuation of $m$. It is important to note 
here that 
one can use the suitable extrapolation scheme ({\it e.g.} ratio method) to 
find the reduced free energy per base pair from the relation
$G(\omega,u)= \lim_{N \rightarrow \infty} \frac{1}{N} \log Z_N(\omega,u)$
and corresponding transition points of the F-T diagram in the 
thermodynamic limit also. However, in our calculation we shall confine 
ourselves to canonical ensemble and set $\epsilon =-1$ in calculating 
all the relevant quantities.

\begin{figure}[t]
\includegraphics[width=3.in]{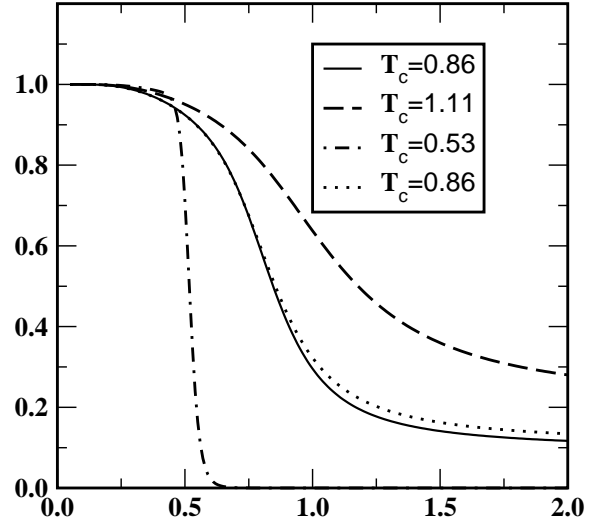}
\caption{Melting profiles of dsDNA under different constrain imposed on the 
ends of the chain. The solid line corresponds to the Figure 4(a). 
The dashed line represents the case illustrated in Fig. 4(b). 
The dotted line is for the situation shown in Fig. 4(c) and the 
dotted-dashed line is for Fig. 4(d). 
Foll all these cases $T_c$ is found from the fluctuation of $m$ which is
close to the temperature (melting temperature) where half of the base
pairs are opened. }
\label{fig-5}
\end{figure}

\section{Results}

For the finite size chain, the melting profile of dsDNA 
strongly depends on the constraints  imposed on the strands. For example, 
if we keep one end of both strands fixed and other ends free (Fig. 4 (a)), 
the melting temperature is found to be $0.86$. However, if one    
end of both strands of the dsDNA is kept fixed and other ends are tied together 
(Fig. 4(b)), in that case dsDNA melts at $T= 1.11$. The other possibility is
to tie one end of the dsDNA together and keep only one strand of the other 
side of dsDNA fixed (Fig. 4c). In this case melting takes place at T=0.86. Lastly 
we can fix one end (5'-end) of the first strand and opposite ends (5'-end) 
of the other strand (Fig. 4(d)), the melting occurs at $T =  0.53$. 
The variation in melting temperature 
is due to the reduction in entropy solely arising due to the  imposed 
constrain on the ends of the chain. It may be noted here that in case of 
unzipping and slippage, 
such confinements are being generally imposed by the experimental 
setup, and, therefore, the resultant force-temperature diagrams may differ 
accordingly. In the following, we shall discuss the effect of confinement
shown by Figs. 4(a) \& 4(c) for DNA unzipping and slippage respectively.

\subsection{Pulling at 5'-3' end of opposite strands: DNA unzipping}
Pulling at one end of dsDNA (5'-3' end) results DNA unzipping. We keep one 
end of the dsDNA fixed (Fig. 4 (a) and apply a force $f$ on the other 
end as shown in Fig. 1b. The force temperature
diagram shown in Fig. 6  is obtained from the maxima of fluctuation of 
$m$ with T for a given force $f$. 
\vspace {0.2in}
\begin{figure}[h]
\includegraphics[width=3.4in]{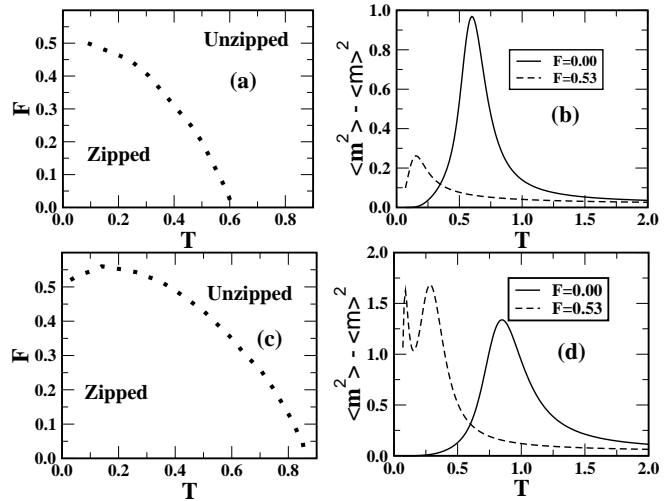}
\caption{Force temperature diagram of DNA unzipping. (a) For the model introduced 
in ref. \cite{kgbhat,eye,jcp}. (b) At low temperature there is only one peak in 
the fluctuation curve which shows the absence of re-entrance in the model studied 
in ref. \cite{kgbhat,eye,jcp}. (c) f-t diagram for the model studied here which shows 
the re-entrance at low temperature. (d) The existence of two peaks is evident 
from the fluctuation curve. 
}
\label{fig-6}
\end{figure}
For the sake of comparison, we also 
provide the result of Ref \cite{kgbhat,eye,jcp} in Fig. 6(a), where base pairing interaction is 
carried out along the bond as shown in Fig.2 (a) \& (b). As pointed out above, 
the diagonal interaction gives rise the ground state entropy of the zipped 
state. As a result, we see two peaks in the fluctuation of $m$ (Fig. 6d) 
which gives rise the re-entrance in the force temperature plane (Fig. 6(c)) 
at low temperature. As stated earlier, it  is  absent in the Fig. 6(a)
\cite{kgbhat,eye,jcp}. 
It should be noted here that the melting temperature for the diagonal interaction 
is much higher because of the large contribution arising due to the ground state
entropy of the zipped state. 
\subsection{Pulling at 5'-5' end or 3'-3' end of opposite strand}

\subsubsection{ Dissociation of two strands}
If pulling is fast enough or the chain is heterogeneous, the two strands 
separate completely without any overlap. In short span of time rebinding 
of bases are not possible and rupture takes 
place at some critical force $f_c$ where two strands dissociate completely.
In order to model such process, we consider all conformations of two MASAWs
as shown in Fig. 4(c) along with the conformations 
where the second chain has shifted one unit (Fig. 3(b)) towards the force 
direction. 
Since pulling is fast, there is no contribution of base pairing in the 
displaced partition function. As a result two walks will be non-interacting and 
impose only the confinement arising due to mutual exclusion. The combined partition 
function can be written as

\begin{figure}[t]
\includegraphics[width=2.4in]{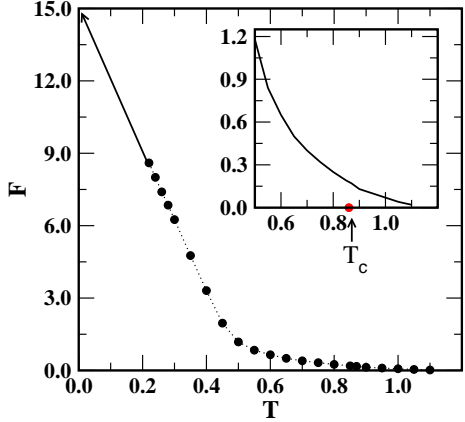}
\caption{The force temperature diagram for the DNA dissociation.
At low temperature force decreases linearly with the temperature.
At $T =0$, it intercepts $y$ axis at $15$ which is the required force
for the rupture. It is clear from the Fig. 5, that above the temperature
$T=0.4$ DNA melts and because of entropic contribution, $F-T$ curve no
more remain linear. Above the melting temperature ($T=0.86$), there are
still some bases are in contact and hence  small force is required for the
complete unbinding as shown in in the inset.
}
\label{fig-7}
\end{figure}

\begin{eqnarray}
Z_N = Z^0 +Z^1
\end{eqnarray}
where $Z^{0}$ is the partition function of the model system in which 
one end of the strand is attached with the AFM tip which may vary in between
0 to N while other end of one strand (Fig. 4(c)) is kept fixed. Here formation 
of base pairing is possible in between ith base of one strand with the ith base
of other strand only. The ground state is complete zipped state. The partition 
function $Z^{1}$ here corresponds to the situation, when one end of 
the second strand has displaced a unit distance in the force direction 
after acquiring complete zipped stretched state i.e $x=N$. 

The force temperature diagram for the rupture is shown in the Fig. 7.  
It is evident from the plot that the nature of diagram is significantly different 
than the DNA unzipping shown in Fig. 6a \& c. This is because in case of 
unzipping, the applied force does not affect the entropy associated with
conformations and bubble while in case of rupture, because of stretching, 
the contribution of entropy goes to zero. Moreover at $f=0$, DNA melts at 
some temperature where half of the bases are still in contact. 
It is evident from the melting profile (Fig. 5) that above the 
melting temperature, there are significant number of base pairs. 
In order to have complete unbinding ({\it i.e.}  no base is in contact), 
one requires still some (vanishingly small) force near the  melting 
temperature as shown in inset of Fig. 7.

The force extension curve obtained in CDE is shown in Fig. 8.  At low 
temperature,
when the dsDNA is in the zipped state, the force brings the dsDNA from coil 
state to the stretched state. Depending on the temperature, at a certain 
critical value of the force, rupture takes place and then the force becomes
zero. The qualitative nature of the force extension curve is similar to the 
one seen in recent experiments \cite{Strunge,Lee_Science94}.
\begin{figure}[t]
\includegraphics[width=2.5in]{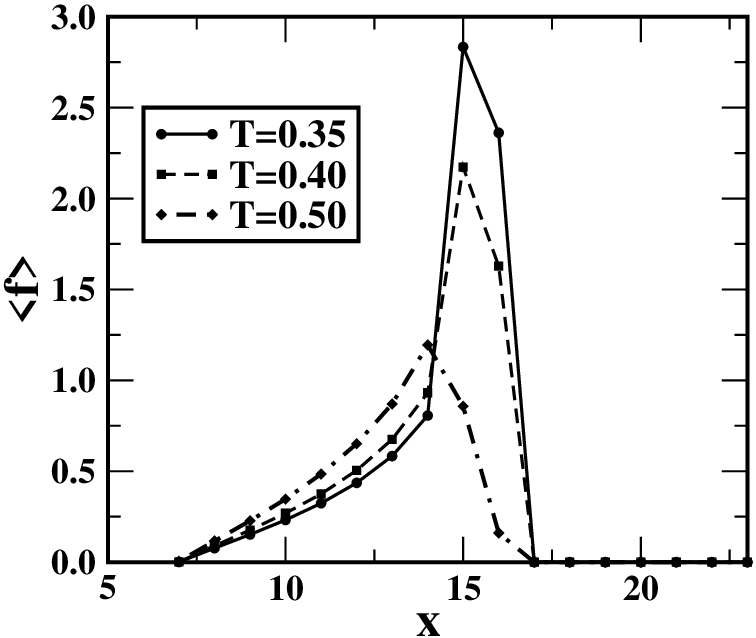}
\caption{The force extension curve in CDE. As temperature increases
unfolding force decreases. The applied force brings system first from 
the coil state to the stretched state and at a critical force 
rupture takes place and force goes to zero as seen in the experiment.
}
\label{fig-8}
\end{figure}
\subsubsection{ Bulge movement}
Due to the formation of a bulge and application of shearing force at 
opposite ends 
of the dsDNA , one strand slowly moves over the other strand along the force
direction. Since pulling is quite slow, there is enough time for unbinding and 
rebinding of the bases. In order to study the effect of bulge on the force 
temperature diagram, we consider the following partition function:
\begin{eqnarray}
Z_N = \sum_{i=0}^{N} Z^{i}
\end{eqnarray}

Unlike the model for rupture, we calculate the partition function
$Z^{1}$, where allowed the formation of the base pairs 
in between ith base of one strand with  the (i+1)th base of the other 
strand when the chain slides one unit distance along the force direction.
Similarly $Z^{2}$ corresponds to the situation where the chain slides two 
units along the force direction and base pairing now takes place in between 
ith base of one strand with  (i+2) base of the other chain and so on.
In quasi static equilibrium this represents the bulge movement along the 
chain.

\begin{figure}[t]
\includegraphics[width=2.5in]{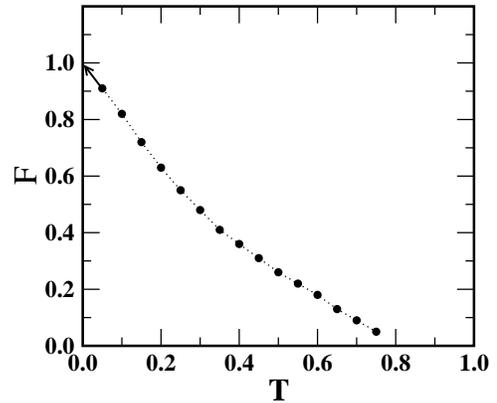}
\caption{The force temperature diagram for the DNA slippage. In this case
DNA separates at much lower force compare to the one shown in Fig. 7.
The other features remain same as of Fig. 7.
}
\label{fig-9}
\end{figure}

The force temperature diagram is shown in Fig. 9. The nature of phase boundary
between zipped state and open  state is different than the one obtained 
for DNA unzipping (Fig. 6(c)) but similar to the dissociation of two strands
(Fig. 7). Moreover, the magnitude of the required force is
much less than the one found for the dissociation of two strands. 
At low temperature, the entropy contribution is negligible and hence 
force required to break a base pairing is nearly equal to 1. However, 
at higher temperature, contribution arises due to entropy, the applied 
force decreases with the temperature.
The force extension curve in CDE ensemble has been shown in Fig. 10. With the 
rise of force, the dsDNA acquires the stretched state. Because of the 
formation of bulges and applied force, chain slides over the other chain 
along the force direction. This is evident from Fig. 10, where without increase
in the applied force, extension increases. It may be noted here that the 
force required to bring chain from coiled state to the stretched state for both
cases rupture and slippage are the same. However, in order to have dissociation, a large
force is required while for the slippage comparatively less force
is needed.

\begin{figure}[h]
\vspace{0.2in}
\includegraphics[width=2.5in]{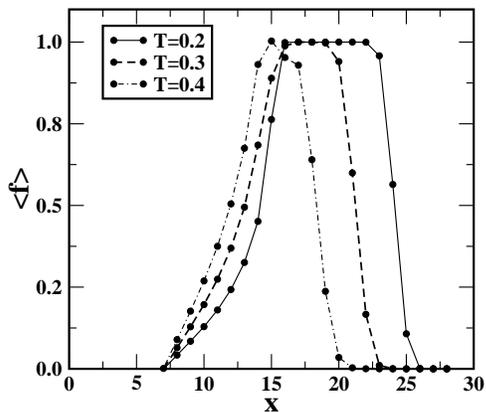}
\caption{The force extension curve in CDE. At low temperature, there
is significant number of overlaps of the base, therefore strand slides
more over the other strand. As temperature increases, number of base
pairs decrease and hence width also decreases. Above certain temperature,
chain dissociates.
}
\label{fig-10}
\end{figure}

\section{Conclusions}
Because of the constrained imposed on the pulling end of dsDNA, there are
significant differences in the melting temperature and melting profile. 
Inclusion of diagonal interaction in the model shows the 
re-entrance in force-temperature diagram of DNA unzipping which were remain
elusive in earlier studies \cite{kgbhat,eye,jcp}. Furthermore, with proper 
modification in the
model we could describe the phenomena like stretching, unzipping, 
dissociation and slippage of dsDNA. The force-temperature diagram of 
slippage and dissociation of dsDNA are significantly different than the 
DNA unzipping. This
is mainly because in dsDNA unzipping, entropy of the chain  and force 
competes with the enthalpy while in dissociation and slippage, it is the
result of an applied force and enthalpy only. Being in stretched state, 
the entropic contribution of the chain is almost zero.

At low temperature, the qualitative nature of the force extension curve 
for dissociation of dsDNA (Fig. 8) is similar to the one observed in 
experiments \cite{Strunge, Lee_Science94}. At high temperature, few bases 
are opened and hence applied force decreases as shown in Fig.7. The 
slippage like transition has already been seen in experiments where the 
existence of plateau has been understood in the form of re-annealing of 
two strands. The qualitative nature of the plateau obtained here is similar 
to one seen in experiment \cite{gaub}. 

At $T = 0$, Eq. 6 gives the force required for rupture which is 
equal to N. This is  evident from the Fig. 7. Up to $T= 0.45$, the number
of intact 
base remain ($N'$) equal to $N$ and hence entropy associated with open state
is zero. The value found from the above equation matches exactly with the 
one shown in Fig. 7 up to $T= 0.45$. Above this temperature,  $N'$ decreases 
with 
temperature and hence bubble forms. Therefore, more force is needed to 
keep system in the stretched state. This is reflected in Fig. 7 where the 
phase boundary between zipped and open states bends. For slippage like 
transition, $x = N$ and hence required force is equal to $1$ consistent 
with the plot shown in Fig. 9.

In this paper, we consider effective base 
pairing energy and hence interaction associated with inter strand and intra-strand stacking 
interaction have been ignored. If one also includes these interactions in the model 
introduced in Ref \cite{eye}, one may get different response for pulling 
at 5'-5' end 3'-3' end \cite{Cludia}. At this 
stage of time a long chain simulation along with hetero sequence is 
needed to understand the mechanism of slippage and dissociation at
vanishingly small force.

\section{Acknowledgment}
We would like to acknowledge the financial supports from the Department of
Science and Technology, India and generous computer resource of MPIPKS, 
Dresden, Germany.


\end{document}